# Two FPGA Case Studies Comparing High Level Synthesis and Manual HDL for HEP applications

Marc-André Tétrault, *Member, IEEE*

*Abstract*—Real time data acquisition systems in nuclear science often rely on high-speed logic designs to reach the fast data rate requirements. They are mostly coded in a hardware description language (HDL). However, in recent years, high level synthesis (HLS) compilers have appeared, with the notable advantage that they rely on the widespread C/C++ syntax. This paper's aim is to outline differences between HDL and C/C++ HLS based designs for two real-time data acquisition modules used in nuclear science. The first module is a real-time crystal identification module, and the second is a compact event timestamp sorting module. This evaluation was done by an experienced VHDL programmer with no prior HLS training.

For the crystal identification module, both HDL and HLS versions have the same event processing interval, and the HLS implementation consumes twice as many lookup tables and flip flops as the HDL version. On the other hand, the HLS version took half the time to write and debug. For the sorter module, the HLS version requires about 3 to 4 times more logic resources, with a slightly longer processing interval. It was also completed in half the time compared to the original HDL code. While different compiler directives can still be explored to improve source code clarity, resource usage and timing closure in these designs, this trial shows that HLS is a compelling alternative to custom HDL implementations for real time systems in nuclear and plasma science.

## I. INTRODUCTION

DATA acquisition systems for high energy physics experiments and nuclear medical instrumentation handle very large amounts of data, often from thousands [1] or millions [2] of channels. These systems require real time, digital data processing modules to quickly extract and select the significant portions of the data stream. Depending on the complexity, they are either implemented in computing platforms, field programmable gate arrays (FPGA) or application-specific integrated circuits (ASIC).

Implementing FPGA or ASIC high-speed digital designs have long relied on hardware description languages (HDL, Verilog or VHDL), where designers have good control on how the synthesis software translates it into actual logic circuits. Labview and Matlab, among others, have added HDL generators to their proprietary software suites, allowing a higher level of abstraction. More recently, high level synthesis (HLS) provides the same advantage but using the C/C++ generic language, with compilers from at least five major vendors (Xilinx, Altera/Intel, Cadence, Synopsys and Mentor). Some case studies show that although HLS does not reach manual HDL performance, it can come quite close but with a much shorter design time [3], [4].

This work aims to provide two different case studies from designs previously used in real time nuclear medical imaging, rewritten for HLS by an experienced VHDL programmer with no prior training with HLS compilers. The goal is to give perspective to the community on the benefits and limitations of the HLS methodology in relevant high energy physics context, with focus on initial contact with the new tool set.

## II. MATERIALS AND METHODS

### A. Cristal Identification Analysis Module

The first reference design is a crystal identification module written in 2008, whose real time implementation was previously published in a firmware update for the LabPET scanner [5]. It is comprised of 6 major steps: 1- baseline estimation and correction, 2- maximum value search with interpolation, 3- normalization, 4- phase identification (through interpolation), 5- Wiener filter multiply-accumulate bloc and 6- Wiener filter matrix inversion. The original design flow used an initial high-level Matlab model, followed by a hardware-friendly, fixed-point Matlab intermediate model from which was derived a VHDL implementation for the Xilinx Virtex-II Pro. Real-world experimental data was used to exercise the VHDL and its results are bit-accurate with the fixed-point intermediate model. The original source code, compiler and synthesis reports are readily available, and extensive logbook entries and version control history provide half-day accuracy on design time span.

The C++ HLS version was designed in a similar manner: an initial double-precision high-level analysis code, followed by a HLS aware code which can switch between floating point and fixed point operations through C templates. The high-level model was not directly used as some changes are required to ensure the compiler understands the data flow nature of the module and infers a coherent implementation. The real-world

Uploaded to arXiv on June 20th 2018. This work was supported in part by the Natural Sciences and Engineering Research Council of Canada (NSERC), by the National Institute of Health under grant P41EB022544, and by CMC Microsystems.

M.-A. Tétrault is with the Massachusetts General Hospital and Harvard Medical School, Boston, MA, 02129, USA (mtetrault@mgh.harvard.edu).

experimental data was used again to validate the design. The code was synthesized using Xilinx Vivado HLS 2017.2, targeting the Zynq-7000 family. The design time span was annotated in logbook entries and version control history.

The comparison scheme between the two methods includes design time, minimum clock period from the synthesis tool and logic resource consumption: RAM blocks, multiplier block, flip flops and lookup tables (LUT). To more easily compare over the 5-generation gap in FPGA technology, the original VHDL was also resynthesized for the Zynq-7000 FPGA without any code changes.

*B. Front-End Sorter*

The second reference design is a real time front-end event sorter used in a distributed coincidence detection engine (Figure 3 in [6]). It uses a memory block as an iterative shift register and inserts timestamps one by one in chronologically correct locations relative to all other events already in the memory. While moderately fast, it has a very small footprint, good average throughput and high operational frequency, making it an efficient and essential part of the coincidence engine. An updated HDL version compared to [6] is used here for reference, version which includes additional pipeline stages to reach maximum operational frequency in recent Xilinx FPGA generations.

The module was rewritten in HLS-aware C code, synthesized using the Xilinx Vivado HLS 2017.2 and targeting the Zynq-7000 family. Comparison metrics include logic resource usage, minimal clock period and event interval. In this module, the event interval is directly related to the sorting array size and any additional pipeline stages, and so to establish a baseline the memory depth was fixed here to 100 events.

III. RESULTS

*A. Crystal Identification Analysis Module*

Table I reports on the logic resource usage and performance of the VHDL and HLS versions of the crystal identification module. The HLS version requires the same amount of memory blocks, a few more multipliers, but more than twice the flip flops and lookup tables. HLS latency is higher than the VHDL counterpart, but this is not a drawback for the intended application. More importantly, the event interval (related to the event rate) remains the same, whose lowest value is limited by the event data length (36 16-bit words), ensuring maximum throughput.

The original design was created in two phases. The first phase took about 15 days and includes writing the high level Matlab model, the hardware-friendly model and evaluating the fixed point dynamic range to balance performance and logic usage. The second phase took 14 days and includes writing the VHDL, writing its test bench and ensuring bit-accurate equivalence with the Matlab fixed-point model. The total is therefore about 29 work days.

It took about 3 days to convert both Matlab models into C and hardware friendly C code. However, this value is inappropriate for the comparison as the design issues were already solved. Since Matlab and C are very similar, the first phase should be presumed to be the same 15 days rather than 3. The second half, HLS synthesis, first required adding one-way data streaming behavior to functions, which generally meant adding an array argument to functions and a copy operation in the *for* loops. The HLS tool then required about a dozen simple directives and a few design iterations to correct user mistakes to obtain the result shown in Table I. Additionally, there is no separate test bench between the high-level models and the HDL, saving additional time. It took about 5 days for these steps, but included exploratory self-training, as well as trial/error iterations to understand the proper compiler directives. For an experienced HLS designer, this is estimated to about 1 to 3 days of work, some steps of which can be done during the hardware-friendly code adjustments.

Lastly, the clock period for HLS is also significantly longer than the VHDL version. The slowest unit here is in the parallel multiply-accumulate stage of the Wiener filter. Other arithmetically intensive blocks (interpolation) report 8 or 9 ns clock period, while simple control blocks are below 7 ns. Further exploration should provide better results once the tool is better understood, but it is not clear if it could reach the same 5.3 ns obtained with the HDL module without increasing the event interval.

Table I – Crystal identification design outcomes

|  | Manual HDL | | HLS |
|---|---|---|---|
|  | Virtex-II Pro | Zynq 7000 | Zynq 7000 |
| Clock period | 6.78 ns | 5.30 ns | 10.09 ns |
| Flip Flops | 1762 | 1736 | 3865 |
| LUT | 1816 | 1438 | 3453 |
| RAM Blocks | 3 | 3 | 3 |
| Multipliers | 23 | 23 | 25 |
| Event Interval | 36 clocks | 36 clocks | 36 clocks |
| Latency | 252 clocks | 252 clocks | 294 clocks |

*B. Frond-End Sorter*

Table II reports on the synthesis and operational results for the sorter module with a memory depth of 100 events, and two different target clock periods for the HLS version. Only the clock constraint was changed, the source code was identical in both cases. Here as well, the HLS design has an overall higher resource count. Another notable difference is the sorting interval, which requires 8 extra clock cycles in the fast HLS version due to additional pipelining. This additional delay introduces a minor but non-negligible performance penalty. The design time difference is again about a factor of two, with 5 days for the original VHDL and 2 days for the HLS.

Table II – Sorter design outcomes

|  | Manual HDL | HLS | |
|---|---|---|---|
|  | Zynq 7000 | Zynq 7000 | Zynq 7000 |
| Clock period | 3.3 ns | 10 ns | 3.3 ns |
| Flip Flops | 187 | 150 | 522 |
| LUT | 75 | 351 | 471 |
| RAM Blocks | 1 | 2 | 2 |
| Interval | 100 clocks | 104 clocks | 108 clocks |

## IV. Discussion

The crystal identification module is a good example where the data streams through successive signal processing functions. The unidirectional data flow made it very straightforward to adapt the natural C code to HLS, and very few special compiler directives are required to generate something close to the original VHDL behavior. Most HLS-specific constructs will seem somewhat superfluous to mainstream C++ programmers, but generally do not obstruct code clarity. On the other hand, high performance sections can require a peculiar coding style to help the compiler infer the correct hardware, and might conceal the designer's intent. Because of this, C programmers without FPGA experience will at first likely have a hard time obtaining optimal results or understand hardware friendly coding styles. The situation is similar to DSP programming, where some devices have parallel multiply-accumulate hardware but need specific coding style or compiler directives to exploit them. Good understanding of the underlying hardware remains a key element to reach compact or high-performance circuits. On the other hand, once experienced FPGA programmers gain HLS experience, they should save significant amounts of time spent on writing and debugging HDL models.

Compared with crystal identification, the sorter module required a significantly unintuitive C code to reach performance near-equivalence with the original VHDL implementation. Indeed, the original VHDL exploits specific hardware and situational optimizations coupled to a continuous circular memory operation to reach an optimal performance. At this document's publishing time, the only HLS path found so far to reach near-equivalence with VHDL is to explicitly define the pipeline stages in the HLS C code. The code is therefore 50% longer than a natural and intuitive representation of the sorting sequence. Nonetheless, the HLS code was still faster to write and debug than the VHDL equivalent. Furthermore, HLS should ease code portability towards future FPGA generations. For example, it took 5 days to convert the original Virtex-II Pro VHDL code to Virtex-5, from 100 MHz operation to 300 MHz. The changes in the memory primitives required rescheduling the logic. On the other hand, the HLS compiler automatically adjusted the design to match the clock constraint and inserted additional stages as required without any changes to the original code.

## V. Conclusion

High level synthesis (HLS) provides a new entry method for designers, using the C/C++ language for a higher level of abstraction than the usual hardware description languages. As shown in this paper, high energy physics experiments can benefit from the fast design cycle of this methodology, the quick turnaround of full updates or complete changes in a processing module. The increased logic footprint might be a concern, but the large resource availability in FPGA devices available today may attenuate this issue. The designers still must understand the underlying hardware and sequential logic design principles, which means that traditional C code requires some modifications to obtain optimal results. For dataflow based designs such as those used in high energy physics, the HLS methodology is an appealing addition to the design toolbox.